# Utsu aftershock productivity law explained from geometric operations on the permanent static stress field of mainshocks

Arnaud Mignan

*Uploaded 20 June 2017*

Institute of Geophysics, Swiss Federal Institute of Technology, Zurich

*Correspondence to:* arnaud.mignan@sed.ethz.ch

*Abstract:* The aftershock productivity law, first described by Utsu in 1970, is an exponential function of the form $n \propto \exp(\alpha M)$ where $n$ is the number of aftershocks, $M$ the mainshock magnitude, and $\alpha$ the productivity parameter. The Utsu law remains empirical in nature although it has also been retrieved in static stress simulations. Here, we explain this law based on Solid Seismicity, a geometrical theory of seismicity where seismicity patterns are described by mathematical expressions obtained from geometric operations on a permanent static stress field. We recover the form $n \propto \exp(\alpha M)$ but with a break in scaling predicted between small and large magnitudes $M$, with $\alpha = 1.5\ln(10)$ and $\ln(10)$, respectively, in agreement with results from previous static stress simulations. We suggest that the lack of break in scaling observed in seismicity catalogues (with $\alpha \approx \ln(10)$) could be an artefact from existing aftershock selection methods, which assume a continuous behavior over the full magnitude range. While the possibility for such an artefact is verified in simulations, the existence of the theoretical kink remains to be proven.

**Introduction**

Aftershocks, the most robust patterns observed in seismicity, are characterized by three empirical laws, which are functions of time (e.g., Utsu et al., 1995; Mignan, 2015), space (e.g., Richards-Dinger et al., 2010) and mainshock magnitude (Utsu, 1970a; b; Ogata, 1988). The present study focuses on the latter relationship, i.e., the Utsu aftershock productivity law, which describes the total number of aftershocks $K$ produced by a mainshock of magnitude $M$ as

$$K(M) = K_0 \exp[\alpha(M - m_0)] \tag{1}$$

with $m_0$ the minimum magnitude cutoff (Utsu, 1970b; Ogata, 1988). This relationship was originally proposed by Utsu (1970a; b) by combining two other empirical laws, the Gutenberg-Richter relationship (Gutenberg and Richter, 1944) and Båth's law (Båth, 1964), respectively:

$$\begin{cases} N(\geq m) = A\exp[-\beta(m - m_0)] \\ \phantom{x}N(\geq M - \Delta m_B) = 1 \end{cases} \tag{2}$$

with β the magnitude size ratio (or $b = \beta/\ln(10)$ in base-10 logarithmic scale) and $\Delta m_B$ the magnitude difference between the mainshock and its largest aftershock, such that

$$K(M) = N(\geq m_0|M) = \exp(-\beta \Delta m_B)\exp[\beta(M - m_0)] \tag{3}$$

with $K_0 = \exp(-\beta \Delta m_B)$ and $\alpha = \beta$. Eq. (3) was only implicit in Utsu (1970a) and not exploited in Utsu (1970b) where $K_0$ was fitted independently of the value taken by Båth's parameter $\Delta m_B$. The α-value was in turn decoupled from the β–value in later studies (e.g., Seif et al. (2017) and references therein).

Although it seems obvious that Eq. (1) can be explained geometrically if the volume of the aftershock zone is correlated to the mainshock surface area $S$ with

$$S(M) = 10^{M-4} = \exp[\ln(10)(M - 4)] \tag{4}$$

(see physical explanation of Eq. (4) in Kanamori and Anderson, 1975), there is so far no analytical, physical expression of Eq. (1) available. Although Hainzl et al. (2010)

retrieved the exponential behavior in numerical simulations where aftershocks were produced by the permanent static stress field of mainshocks of different magnitudes, it remains unclear how $K_0$ and $\alpha$ relate to the underlying physical parameters.

The aim of the present article is to explain the Utsu aftershock productivity equation (Eq. 1) by applying a geometrical theory of seismicity (or "Solid Seismicity"), which has already been shown to effectively explain other empirical laws of both natural and induced seismicity from simple geometric operations on a permanent static stress field (Mignan, 2012; 2016). The theory is applied here for the first time to the case of aftershocks.

**Physical Expression of Aftershock Productivity**

"Solid Seismicity", a geometrical theory of seismicity, is based on the following Postulate (Mignan et al., 2007; Mignan, 2008, 2012; 2016):

**Solid Seismicity Postulate (SSP):** *Seismicity can be strictly categorized into three regimes of constant spatiotemporal densities – background $\delta_0$, quiescence $\delta_-$ and activation $\delta_+$ (with $\delta_- \ll \delta_0 \ll \delta_+$) - occurring respective to the static stress step function:*

$$\delta(\sigma) = \begin{cases} \delta_- & , \sigma < -\Delta o_* \\ \delta_0 & , \sigma \leq |\pm \Delta o_*| \\ \delta_+ & , \sigma > \Delta o_* \end{cases} \qquad (5)$$

*with $\Delta o_*$ the background stress amplitude range.*

Based on this Postulate, Mignan (2012) demonstrated the power-law behavior of precursory seismicity in agreement with the observed time-to-failure equation (Varnes, 1989), while Mignan (2016) demonstrated both the observed parabolic spatiotemporal front and the linear relationship with injection-flow-rate of induced seismicity (Shapiro and Dinske, 2009). It remains unclear whether the SSP has a physical origin or not. If not, it would still represent a reasonable approximation of the linear relationship between event production and static stress field in a simple clock-change model (Hainzl et al., 2010) (Fig. 1a). The power of Eq. (5) is that it allows defining seismicity patterns in terms of "solids" described by the spatial envelope $r_* = r(\sigma = \pm \Delta o_*)$. The spatiotemporal rate of seismicity is then a mathematical expression defined by the density of events δ times the volume characterized by $r_*$ (see previous demonstrations in Mignan et al., 2007; Mignan, 2011; 2012; 2016 where simple algebraic expressions were obtained).

In the case of aftershocks, we define the static stress field of the mainshock by

$$\sigma(r) = -\Delta\sigma_0 \left[ \left(1 - \frac{c^3}{(r+c)^3}\right)^{-1/2} - 1 \right] \quad (6)$$

with $\Delta\sigma_0 < 0$ the mainshock stress drop, $c$ the crack radius and $r$ the distance from the crack. Eq (6) is a simplified representation of stress change from slip on a planar surface in a homogeneous elastic medium. It takes into account both the square root singularity at crack tip and the $1/r^3$ falloff at higher distances (Dieterich, 1994) (Fig. 1b). It should be noted that this radial static stress field does not represent the geometric complexity of Coulomb stress fields (Fig. 2a). However we are here only interested in the general behavior of aftershocks with Eq. (6) retaining the first-order characteristics of this field (i.e., on-fault seismicity; Fig. 2b), which corresponds to the case where the mainshock relieves most of the regional stresses and aftershocks occur on optimally oriented faults. It is also in agreement with observations, most aftershocks being located on and around the mainshock fault traces in Southern California (Fig. 2c; see section "Observations & Model Fitting"). The occasional cases where aftershocks occur off-fault (e.g., Ross et al., 2017) can be explained by the mainshock not relieving all of the regional stress (King et al., 1994) (Fig. 2d).

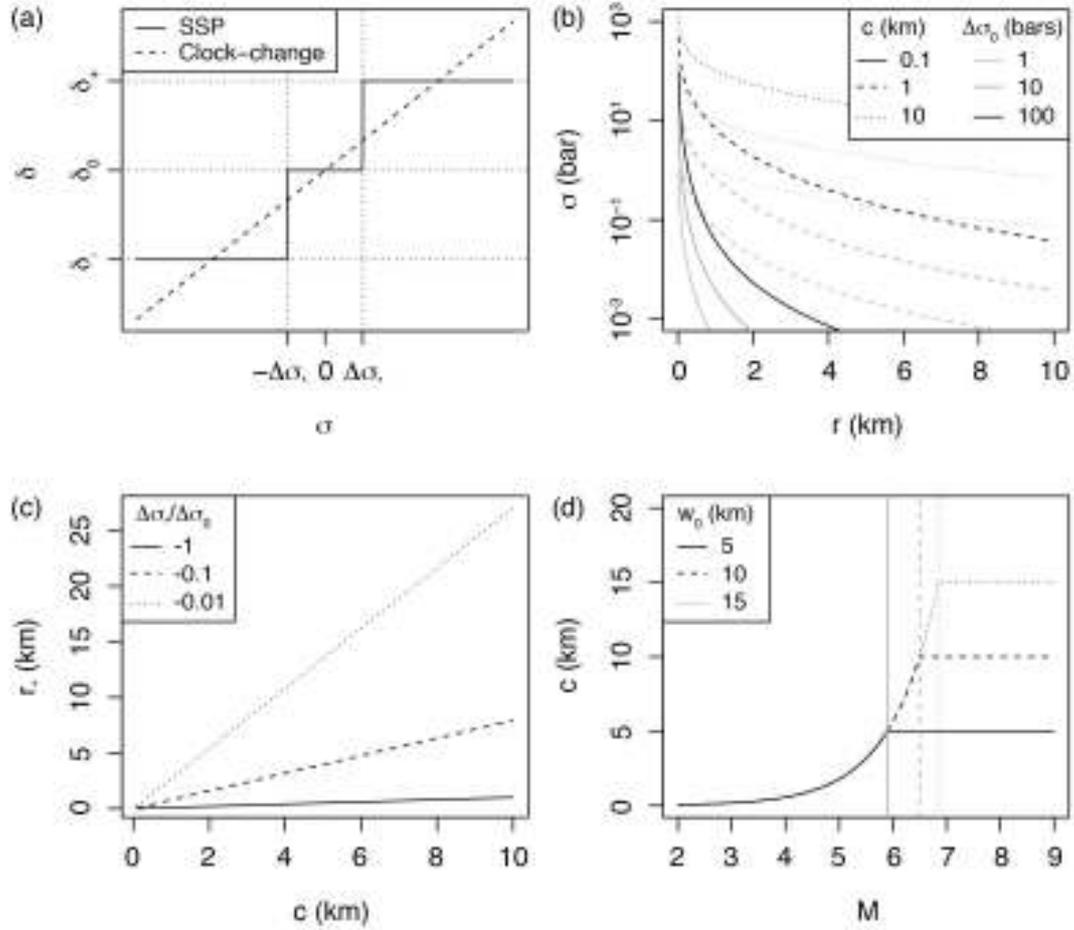

**Figure 1.** Definition of the aftershock solid envelope in a permanent static stress field: (a) Event density stress step-function $\delta(\sigma)$ (Eq. 5) of the Solid Seismicity Postulate (SSP) in comparison to the linear clock-change model; (b) Static stress $\sigma$ versus distance $r$ for different effective crack radii $c$ and rupture stress drops $\Delta\sigma_0$ (Eq. 6); (c) Linear relationship between effective crack radius $c$ and aftershock solid envelope radius $r_*$ for different $\Delta\sigma_*/\Delta\sigma_0$ ratios (Eq. 7); (d) Relationship between mainshock magnitude $M$ and effective crack radius $c$ for different seismogenic widths $w_0$ (Eq. 8).

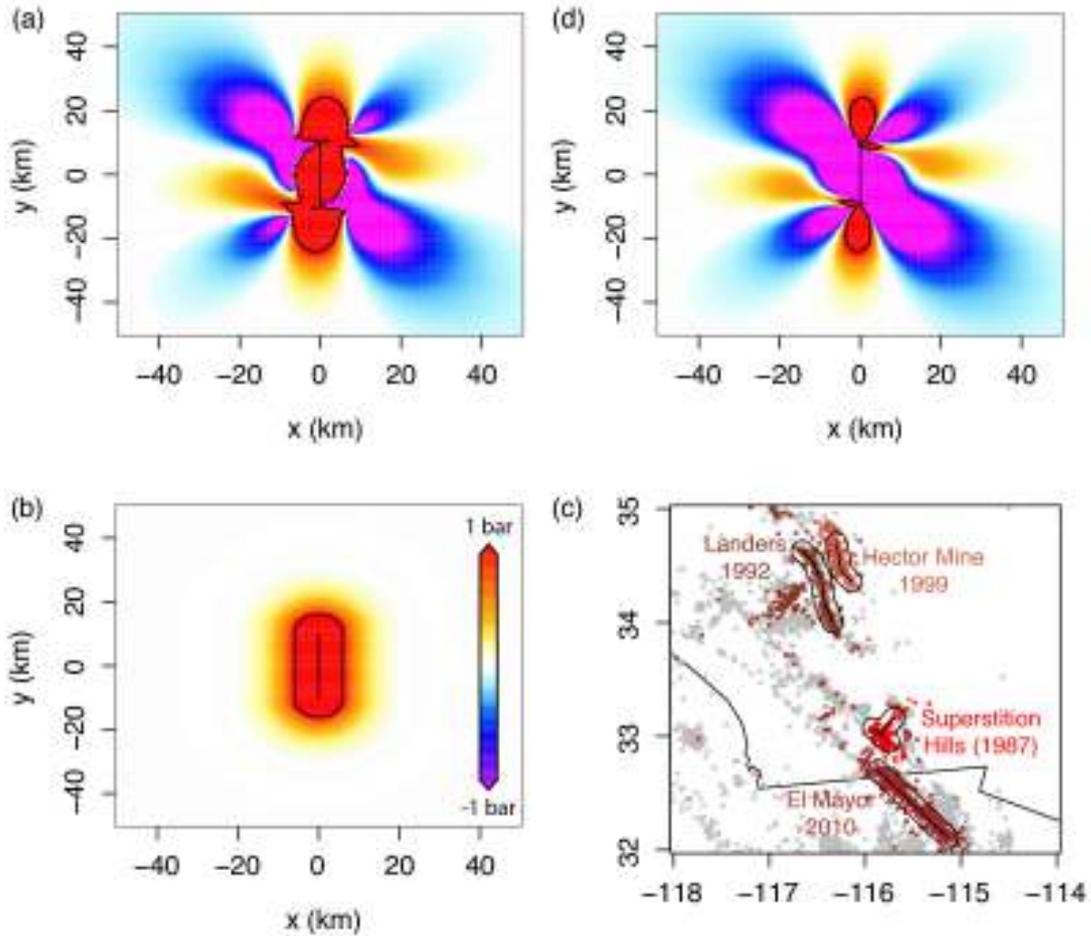

**Figure 2.** Possible static stress fields and inferred aftershock spatial distribution: (a) Right-lateral Coulomb stress field for optimally oriented faults, where the mainshock relieves all of the regional stresses $\sigma_r$ = 10 bar, with $\Delta\sigma_0 \approx -Gs/L \approx$ - 10 bar ($G = 3.3.10^5$ bar the shear modulus, $s$ = 0.6 m the slip, $L$ = 20 km the fault length, and $w$ = 10 km the fault width); (b) Radial static stress field computed from Eq. (6) with $\Delta\sigma_0$ = -10 bar and $c = \sqrt{(Lw)/\pi}$ for consistency with (a); (c) Aftershock distribution of the largest strike-slip events in the Southern California relocated catalog, identified here as all events occurring within one day of the mainshock; (d) Right-lateral Coulomb stress field for optimally oriented faults, where the mainshock relieves only a fraction of the regional stresses $\sigma_r$ = 100 bar with $\Delta\sigma_0$ = -10 bar (same rupture as in (a)) – The black contour represents 1 bar in (a), (b) and (d), and a 10 km distance from rupture in (c). Coulomb stress fields of (a) and (d) were computed using the Coulomb 3 software (Lin and Stein, 2004; Toda et al., 2005).

For $r_* = r(\sigma = \Delta o_*)$, Eq. (6) yields the aftershock solid envelope of the form:

$$r_*(c) = \left\{ \frac{1}{\left[1-\left(1-\frac{\Delta\sigma_*}{\Delta\sigma_0}\right)^{-2}\right]^{1/3}} - 1 \right\} c = Fc \tag{7}$$

, function of the crack radius $c$ and of the ratio between background stress amplitude range $\Delta o_*$ and stress drop $\Delta\sigma_0$ (Fig. 1c). With $\Delta\sigma_0$ independent of earthquake size (Kanamori and Anderson, 1975; Abercrombie and Leary, 1993) and $\Delta o_*$ assumed constant, $r_*$ is directly proportional to $c$ with proportionality constant, or stress factor, $F$ (Eq. 7). Geometrical constraints due to the seismogenic layer width $w_0$ then yield

$$c(M) = \begin{cases} \left(\frac{S(M)}{\pi}\right)^{1/2} & , S(M) \leq \pi w_0^2 \\ w_0 & , S(M) > \pi w_0^2 \end{cases} \tag{8}$$

with $S$ the rupture surface defined by Eq. (4) and $c$ becoming an effective crack radius (Kanamori and Anderson, 1975; Fig. 1d). Note that the factor of 2 (i.e., using $w_0$ instead of $w_0/2$) comes from the free surface effect (e.g., Kanamori and Anderson, 1975; Shaw and Scholz, 2001).

The aftershock productivity $K(M)$ is then the activation density $\delta_+$ times the volume $V_*(M)$ of the aftershock solid. For the case in which the mainshock relieves most of the regional stress, stresses are increased all around the rupture (King et al., 1994), which is topologically identical to stresses increasing radially from the rupture plane (Fig. 2a-b). It follows that the aftershock solid can be represented by a volume of contour $r_*(M)$ from the rupture plane geometric primitive, i.e., a disk or a rectangle, for small and large mainshocks respectively. This is illustrated in Figure 3a-b and can be generalized by

$$V_*(M) = 2r_*(M)S(M) + \frac{\pi}{2}r_*^2(M)d \tag{9}$$

where $d$ is the distance travelled around the geometric primitive by the geometric centroid of the semi-circle of radius $r_*(M)$ (i.e., Pappus's Centroid Theorem), or

$$d = \begin{cases} 2\pi \left( c(M) + \frac{4}{3\pi} r_*(M) \right) & , c(M) + r_*(M) \leq \frac{w_0}{2} \\ 2w_0 & , c(M) + r_*(M) > \frac{w_0}{2} \end{cases} \quad (10)$$

For the disk, the volume (Eq. 9) corresponds to the sum of a cylinder of radius $c(M)$ and height $2r_*(M)$ (first term) and of half a torus of major radius $c(M)$ and minus radius $r_*(M)$ (second term). For the rectangle, the volume is the sum of a cuboid of length $l(M)$ (i.e., rupture length), width $w_0$ and height $2r_*(M)$ (first term) and of a cylinder of radius $r_*(M)$ and height $w_0$ (second term; see red and orange volumes, respectively, in Figure 3a-c). Finally inserting Eqs. (7), (8) and (10) into (9), we obtain

$$K(M) = \delta_+(m_0) \begin{cases} \left[ \frac{2F}{\sqrt{\pi}} + F^2 \sqrt{\pi} \left( 1 + \frac{4}{3\pi} F \right) \right] S^{3/2}(M) & , S(M) \leq \left( \frac{w_0 \sqrt{\pi}}{2(1+F)} \right)^2 \\ \frac{2F}{\sqrt{\pi}} S^{3/2}(M) + F^2 w_0 S(M) & , \left( \frac{w_0 \sqrt{\pi}}{2(1+F)} \right)^2 < S(M) \leq \pi w_0^2 \\ 2F w_0 S(M) + \pi F^2 w_0^3 & , S(M) > \pi w_0^2 \end{cases}$$

(11)

which is represented in Figure 3d. Considering the two main regimes only (small versus large mainshocks) and inserting Eq. (4) into (11), we get

$$K(M) = \delta_+(m_0) \begin{cases} \left[ \frac{2F}{\sqrt{\pi}} + F^2 \sqrt{\pi} \left( 1 + \frac{4}{3\pi} F \right) \right] \exp\left[ \frac{3\ln(10)}{2} (M-4) \right] & , \text{small } M \\ 2F w_0 \exp[\ln(10)(M-4)] + \pi F^2 w_0^3 & , \text{large } M \end{cases}$$

(12)

which is a closed-form expression of the same form as the original Utsu productivity law (Eq. 1).

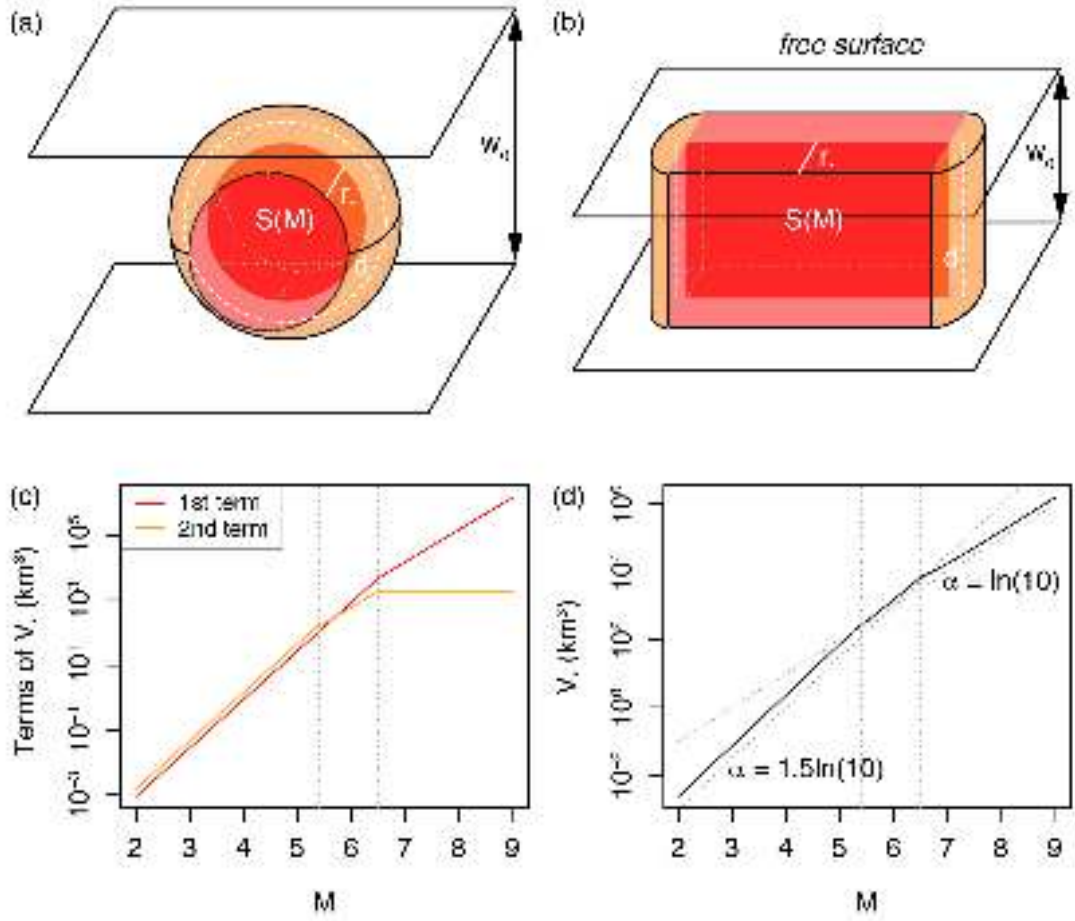

**Figure 3.** Geometric origin of the aftershock productivity law: (a) Sketch of the aftershock solid for a small mainshock rupture represented by a disk; (b) Sketch of the aftershock solid for a large mainshock rupture represented by a rectangle; (c) Relative role of the two terms of Eq. (9), here with $w_0$ = 10 km and $\frac{\Delta\sigma_*}{\Delta\sigma_0}$ = -0.1 (to first estimate c and $r_*$ from Eqs. 8 and 7, respectively); (d) Aftershock productivity law (normalized by $\delta_+$) predicted by Solid Seismicity (Eq. 11). This relationship is of the same form as the Utsu productivity law (Eq. 1) for large $M$ (see text for an explanation of the lack of break in scaling in Eq. 1 for small $M$). Dotted vertical lines represent $M$ for $c(M) + r_*(M) = \frac{w_0}{2}$ and $S(M) = \pi w_0^2$, respectively.

Here, we predict that the α-value decreases from $3\ln(10)/2 \approx 3.45$ to $\ln(10) \approx 2.30$ when switching regime from small to large mainshocks (or from 1.5 to 1 in base-10 logarithmic scale). It should be noted that Hainzl et al. (2010) observed the same break in scaling in static stress transfer simulations, which corroborates our analytical findings. For large *M*, the scaling is fundamentally the same as in Eq. (4). Since that relation also explains the slope of the Gutenberg-Richter law (see physical explanation given by Kanamori and Anderson, 1975), it follows that $\alpha \equiv \beta$, which is also in agreement with the original formulation of Utsu (1970a; b) (Eq. 3).

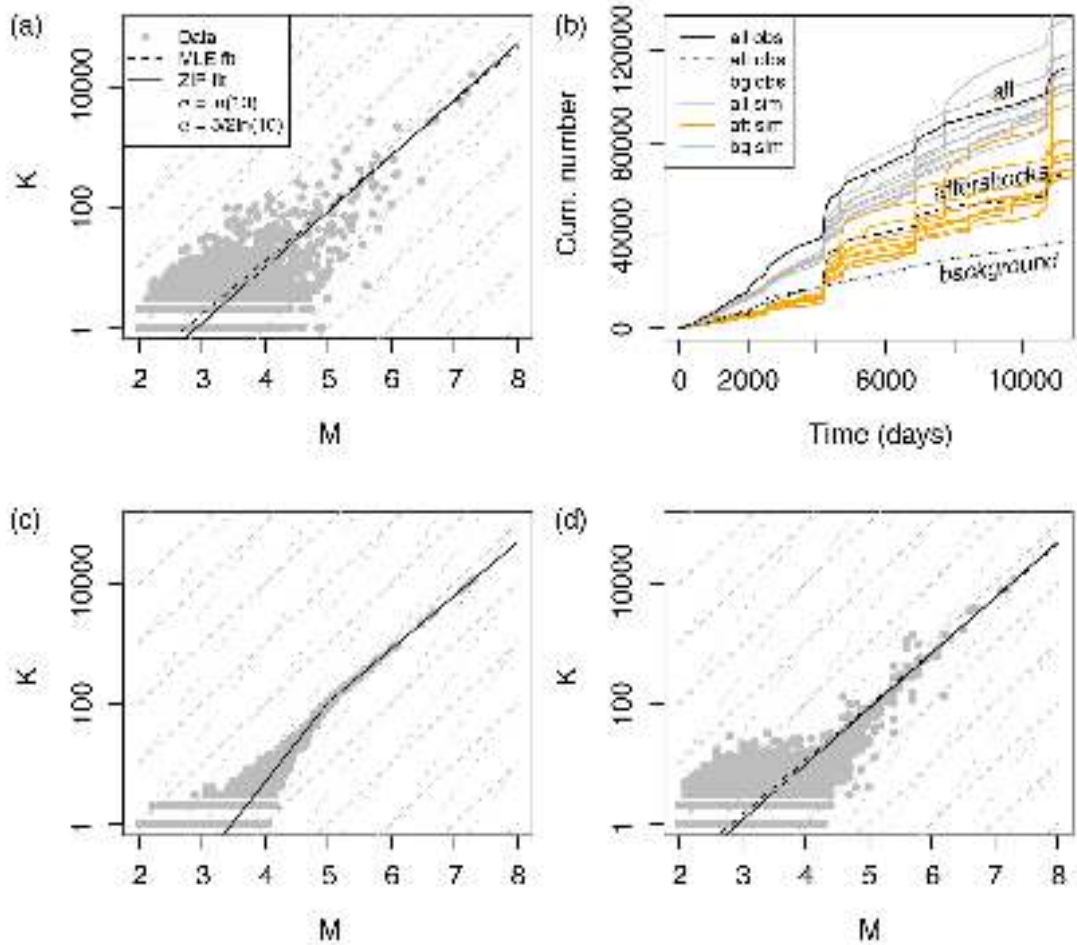

**Figure 4.** Aftershock productivity, observed and simulated, defined as the number of aftershocks $N(m_0 = 2)$ per mainshock of magnitude *M*: (a) Observed aftershock productivity in Southern California with aftershocks selected using the nearest-neighbor method; (b) Seismicity time series with distinction made between background events and aftershocks, observed ("obs", in black, defined from the nearest-neighbor method) and ETAS-simulated ("sim", colored, defined as labeled in ETAS); (c) True simulated aftershock productivity with kink, defined from Eq. (16); (d) Retrieved simulated aftershock productivity with aftershocks selected using the nearest-neighbor method - Data points in (a), (c) and (d) are represented by gray dots; the model fits by Maximum Likelihood Estimation (MLE) method are represented by the dashed and solid black lines for the Poisson and Zero-Inflated Poisson distributions, respectively; dashed and dotted gray lines are visual guides to α = 3/2ln(10) and ln(10), respectively.

**Observations & Model Fitting**

We consider the case of Southern California and extract aftershock sequences from the relocated earthquake catalog of Hauksson et al. (2012) defined over the period 1981-2011, using the nearest-neighbor method (Zaliapin et al., 2008) (used with its standard parameters originally calibrated for Southern California). Only events with magnitudes greater than $m_0$ = 2.0 are considered (a conservative estimate following results of Tormann et al. (2014); saturation effects immediately after the mainshock are negligible when considering entire aftershock sequences; Helmstetter et al., 2005). The observed number of aftershocks *n* produced by a mainshock of magnitude *M* (for a total of *N* mainshocks) is shown in Figure 4a.

We first fit Eq. (1) to the data using the Maximum Likelihood Estimation (MLE) method with the log-likelihood function

$$LL(\theta; X = \{n_i; i = 1, ..., N\}) = \sum_{i=1}^{N} [n_i \ln[K_i(\theta)] - K_i(\theta) - \ln(n_i!)] \qquad (13)$$

for a Poisson process, or, with Eq. (1),

$$LL(\theta = \{K_0, \alpha\}; X) = \ln(K_0) \sum_{i=1}^{N} n_i + \alpha \sum_{i=1}^{N} [n_i(M_i - m_0)] - K_0 \sum_{i=1}^{N} \exp[\alpha(M_i - m_0)] - \sum_{i=1}^{N} \ln(n_i!) \qquad (14)$$

(note that the last term can be set to 0 during *LL* maximization). For Southern California, we obtain $\alpha_{MLE}$ = 2.04 (0.89 in $\log_{10}$ scale) and $K_0$ = 0.23. It should be noted that this approach does not include the case of mainshocks that produce zero aftershock. Therefore we also compute the MLE for the Zero-Inflated Poisson (ZIP) distribution:

$$\begin{cases} \Pr(n_i = 0) = w + (1-w)\exp(-K_i) \\ \Pr(n_i > 0) = (1-w)\frac{K_i^{n_i}}{n_i!}\exp(-K_i) \end{cases} \quad (15)$$

where $w$ is a weighting constant. It finally follows that $\alpha_{MLE(ZIP)} = 2.13$ (0.93 in $\log_{10}$ scale, with $K_0 = 0.15$), corrected for zero-values. This result is in agreement with previous studies in the same region (e.g., Helmstetter et al., 2005; Zaliapin and Ben-Zion, 2013; Seif et al., 2017) and with $\alpha = \ln(10) \approx 2.30$ predicted for large mainshocks in Solid Seismicity. Moreover we find a bulk $\beta_{MLE} = 2.34$ (1.02 in $\log_{10}$ scale) (Aki, 1965), in agreement with $\alpha = \beta$. It should be noted that no significant difference is obtained when computing $\beta_{MLE}$ for background events or aftershocks alone, with $\beta_{MLE} = 2.29$ and 2.35, respectively (0.99 and 1.02 in $\log_{10}$ scale).

We also tested the following piecewise model to identify any break in scaling, as predicted by Eq. (12):

$$K(M) = \begin{cases} K_0 \frac{\exp[\ln(10)(M_{break}-m_0)]}{\exp[\frac{3}{2}\ln(10)(M_{break}-m_0)]} \exp\left[\frac{3}{2}\ln(10)(M-m_0)\right] &, M \leq M_{break} \\ K_0 \exp[\ln(10)(M-m_0)] &, M > M_{break} \end{cases}$$

(16)

but with the best MLE result obtained for $M_{break} = m_0$, suggesting no break in scaling in the aftershock productivity data.

**Role of aftershock selection on productivity scaling-break**

We now identify whether the lack of break in scaling in aftershock productivity observed in earthquake catalogues could be an artefact related to the aftershock selection method. We run Epidemic-Type Aftershock Sequence (ETAS) simulations (Ogata, 1988; Ogata and Zhuang, 2006), with the seismicity rate

$$\begin{cases} \lambda(t,x,y) = \mu(t,x,y) + \sum_{i:t_j<t} K(M_i)f(t-t_i)g(x-x_i, y-y_i|M_i) \\ \qquad\qquad f(t) = c^{p-1}(p-1)(t+c)^{-p} \\ g(x,y|M) = \frac{1}{\pi}\left(de^{\gamma(M-m_0)}\right)^{q-1}\left(x^2+y^2+de^{\gamma(M-m_0)}\right)^{-q}(q-1) \end{cases} \quad (17)$$

Aftershock sequences are defined by power laws, both in time and space (for an alternative temporal function, see Mignan, 2015; 2016b). μ is the Southern California background seismicity, as defined by the nearest-neighbor method (with same *t*, *x*, *y* and *m*). We fix the ETAS parameters to θ = {*c* = 0.011 day, *p* = 1.08, *d* = 0.0019 km², *q* = 1.47, γ = 2.01}, following the fitting results of Seif et al. (2017) for the Southern California relocated catalog and $m_0$ = 2 (see their Table 1). However, we define the productivity *K*(*M*) from Eq. (16) with $M_{break}$ = 5, $K_0$ = 0.23, α = 2.04 and β = 2.3. Examples of ETAS simulations are shown in Figure 4b for comparison with the observed Southern California time series. Figure 4c allows us to verify that the simulated aftershock productivity is kinked at $M_{break}$, as defined by Eq. (16).

We then select aftershocks from the ETAS simulations with the nearest-neighbor method. Figure 4d represents the estimated aftershock productivity, which has lost the break in scaling originally implemented in the simulations. This demonstrates that the theoretical break in scaling predicted in the aftershock productivity law can be lost in observations due to an aftershock selection bias, all declustering techniques assuming continuity over the entire magnitude range. While such an artefact is possible, it yet does not prove that the break in scaling exists. The fact that a similar break in scaling was obtained in independent Coulomb stress simulations (Hainzl et al., 2010) however provides high confidence in our results.

**Conclusions**

In the present study, a physical closed-form expression defined from geometric and static stress parameters was proposed (Eq. 12) to explain the empirical Utsu aftershock productivity law (Eq. 1). This demonstration, combined to the previous ones made by the author to explain precursory accelerating seismicity and induced seismicity (Mignan, 2012; 2016), suggests that most empirical laws observed in seismicity populations can be explained by simple geometric operations on a permanent static stress field. Although the Solid Seismicity Postulate (SSP) (Eq. 5) remains to be proven, it is so far a rather convenient and pragmatic assumption to determine the physical parameters that play a first-order role in the behavior of seismicity. It is also complementary to the more common simulations of static stress loading (King and Bowman, 2003) and static stress triggering (Hainzl et al., 2010). Analytic geometry, providing both a visual representation and an analytical expression of the problem at hand (Fig. 3), represents a new approach to try better understanding the behavior of seismicity. Its current limitation in the case of aftershock analysis consists in assuming that the static stress field is radial and described by Eq. (6) (Dieterich, 1994), which is likely only valid for mainshocks relieving most of the regional stresses and with aftershocks occurring on optimally oriented faults (King et al., 1994). More complex, second-order, stress behaviors might explain part of the scattering observed around Eq. (1) (Fig. 4a). Other $\sigma(r)$ formulations could be tested in the future, the only constraint on generating so-called seismicity solids being the use of the postulated static stress step function of Eq. (5) (i.e., the Solid Seismicity Postulate, SSP).